\documentclass[aps,prx,reprint,superscriptaddress,floatfix,longbibliography]{revtex4-1}
\usepackage{lipsum,amsmath,amssymb}
\usepackage{graphicx,color}

\makeatletter
\let\cat@comma@active\@empty
\makeatother
\begin{document}


\title{Multiphase field models for collective cell migration}


\author{D. Wenzel}
\affiliation{Institute of Scientific Computing, Technische Universit\"at Dresden, 01062 Dresden, Germany}
\author{A. Voigt}
\affiliation{Institute of Scientific Computing, Technische Universit\"at Dresden, 01062 Dresden, Germany}
\affiliation{Center for Systems Biology Dresden (CSBD), Pfotenhauerstr. 108, 01307 Dresden, Germany} 
\affiliation{Cluster of Excellence - Physics of Life, TU Dresden, 01062 Dresden, Germany}

\begin{abstract}
Confluent cell monolayers and epithelia tissues show remarkable patterns and correlations in structural arrangements and actively-driven collective flows. We simulate these properties using multiphase field models. The models are based on cell deformations and cell-cell interactions and we investigate the influence of microscopic details to incorporate active forces on emerging phenomena. We compare four different approaches, one in which the activity is determined by a random orientation, one where the activity is related to the deformation of the cells and two models with subcellular details to resolve the mechanochemical interactions underlying cell migration. The models are compared with respect to generic features, such as solid-to-liquid phase transitions, cell shape variability, emerging nematic properties, as well as vorticity correlations and flow patterns in large confluent monolayers and confinements. All results are compared with experimental data for a large variety of cell cultures. The appearing qualitative differences of the models show the importance of microscopic details and provide a route towards predictive simulations of patterns and correlations in cell colonies.        
\end{abstract}

\pacs{?}

\maketitle


\section{Introduction}

The ability of cells to coordinate their motion is essential
for several in vivo processes, such as morphogenesis, regeneration and cancer invasion \cite{Friedl_NRMCB_2009,Rorth_ARCDB_2009,Scarpa_JCB_2016}. To identify the principles that govern collective cell migration in such systems has seen a growing interest in recent years. Experimental investigations on cell monolayers and epithelial tissue of model systems have shown remarkable patterns and correlations in cell migration. These include an unjamming transition between a glassy phase and a fluid phase \cite{Malinverno_NM_2017,Atia_NP_2018}, the spontaneous formation of vortices and topological defects \cite{Saw_N_2017}, as well as the emergence of active turbulent flows \cite{Blanch_PRL_2018}. The emerging phenomena appear to be rather generic.
A fundamental challenge is to understand how this macroscopic behaviour is linked to the properties of individual cells and physical cell-cell interactions, which is the target of a large variety of modeling approaches. These approaches differ by the level of coarse-graining and range from subcellular lattice models \cite{Graner_PRL_1992} and multiphase field models \cite{Nonomura_PLOS_2012,Camley_PNAS_2014,Palmieri_SR_2015,Mueller_PRL_2019,Wenzel_JCP_2019,Loewe_PRL_2020}, to vertex and voronoi models \cite{Nagai_PMB_2001,Staple_EPJE_2010,Fletcher_BPJ_2014,Li_BPJ_2014,Bi_PRX_2016}, particle models \cite{Redner_PRE_2013,Navarro_SM_2015,Aland_BPJ_2015,Prymidis_JCP_2016,Alaimo_NJP_2016} and continuum models on a multicellular scale \cite{Duclos_NP_2018,DellArciprete_NC_2018,Doostmohammadi_NC_2018}. We refer to \cite{Hakim_RPP_2017,Alert_ARCMP_2020,Moure_ACME_2020} for recent reviews. 

We here concentrate on multiphase field models, which allow for cell deformations and detailed cell-cell interactions, as well as subcellular details to resolve the mechanochemical interactions underlying cell migration. Together with efficient numerics and appropriate computing power these models are well suited to model confluent cell structures and have seen various recent contributions \cite{Nonomura_PLOS_2012,Camley_PNAS_2014,Palmieri_SR_2015,Mueller_PRL_2019,Wenzel_JCP_2019,Loewe_PRL_2020}. They all follow the same methodology but differ in detail. 

The goal of this paper is a systematic comparison of these approaches and their linkage with statistical observables of experiments to provide a route towards predictive simulations of patterns and correlations in cell colonies. After introducing the multiphase field models, discussing microscopic differences, and briefly describing the numerical approach enabling large scale simulations, we address solid-to-liquid phase transition, analyse statistics on shape variability of the cells and the ratio of multicellular rosettes, velocity distributions of emerging topological defects, their stress fields as well as defect density and creation rates. We further study vorticity correlations in large confluent monolayers and flow patterns in confinements. All results are compared with experimental data for a large variety of cell cultures. 

\section{Modeling}

We consider two-dimensional phase field variables $\phi_i$, one for each cell. Values of $\phi_i = 1$ and $\phi_i = -1$ denote the interior and the exterior of a cell, respectively. The cell boundary is defined implicitly by the $\phi_i = 0$ level-set. The dynamics for each $\phi_i$ is considered as
\begin{equation}
    \partial_t \phi_i + v_0 (\mathbf{v}_i \cdot \nabla \phi_i) = \Delta \frac{\delta \mathcal{F} }{\delta \phi_i}, \quad i = 1, \ldots, N,
    \label{eq:phi}
\end{equation}
where $N$ denotes the number of cells, $\mathcal{F}$ is a free energy and $\mathbf{v}_i$ a vector field used to incorporate active components, with a self-propulsion strength $v_0$. We here consider conserved dynamics, which ensures constant volume/area of each cell. The proposed models in \cite{Nonomura_PLOS_2012,Palmieri_SR_2015,Mueller_PRL_2019,Loewe_PRL_2020} consider non-conserved dynamics and enforce the volume/area constraint weakly by an additional penalty energy. The free energy $\mathcal{F} = \mathcal{F}_{CH} + \mathcal{F}_{INT} + \ldots$ contains passive contributions, where
\begin{eqnarray}
\mathcal{F}_{CH} &=& \sum_{i=1}^N \frac{1}{Ca}\int_\Omega \frac{\epsilon}{2}\|\nabla\phi_i\|^2 + \frac{1}{\epsilon}W(\phi_i)\,\text{d}\mathbf{x}, \\
\label{eq:IntEnergy}
\mathcal{F}_{INT} &=& \sum_{i=1}^N \frac{1}{In}\int_\Omega B(\phi_i) \sum_{j\neq i} w(d_j)\,\text{d}\mathbf{x},
\end{eqnarray}
with non-dimensional capillary and interaction number, $Ca$ and $In$, respectively. The first is a Cahn-Hilliard energy, with $W(\phi_i) = \frac{1}{4}(\phi_i^2 - 1)^2$ a double-well potential and $\epsilon$ a small parameter determining the width of the diffuse interface. This energy stabilizes the cell interface. For simplicity we here neglect other properties of the cell boundary, e.g., bending forces. In \cite{Marth_JRSI_2015} they are shown to be negligible in the context of cell migration. The second is an interaction energy with $B(\phi_i) = \frac{3}{\epsilon \sqrt{2}} W(\phi_i)$ an approximation of the delta function of the cell boundary and a cell-cell interaction potential 
\begin{equation}
    w(d_j) = \exp{\big(-\frac{d_j^2}{\epsilon^2}\big)}, \mbox{ with } d_j = - \frac{\epsilon}{\sqrt{2}} \ln{\big( \frac{1 + \phi_j}{1 - \phi_j}\big)} \quad
    \label{eq:interaction}
\end{equation}
approximating a short range repulsion potential, with signed distance function $d_j$ computed from the equilibrium $\tanh$-profile of the phase field $\phi_j$, see \cite{Marth_IF_2016,Marth_JFM_2016}. Most previous multiphase field models consider the interaction only effectively using terms proportional to $\phi_i^2 \phi_j^2$ for cell-cell repulsion and $\|\nabla \phi_i\|^2 \|\nabla \phi_j\|^2$ for cell-cell attraction. The approach in eq. (\ref{eq:interaction}) offers the possibility to also consider more realistic potentials. Most significantly the proposed models \cite{Nonomura_PLOS_2012,Camley_PNAS_2014,Palmieri_SR_2015,Mueller_PRL_2019,Wenzel_JCP_2019,Loewe_PRL_2020} differ in the self-propulsion term, the definition of $\mathbf{v}_i$ in eq. (\ref{eq:phi}). 

In \cite{Loewe_PRL_2020} the propulsion speed is the same for each cell, but the direction of motion, determined by the angle $\theta_i$ is controlled by rotational noise $d \theta_i(t) = \sqrt{2 D_r} d W_i(t)$, with diffusivity $D_r$ and a Wiener process $W_i$. With $\mathbf{v}_i^{ran} = (\cos{\theta_i}, \sin{\theta_i})$ the governing equations can be viewed as a generalization of a model for active Brownian particles \cite{Fily_PRL_2012,Redner_PRE_2013,Wysocki_EPL_2014} to one for a system of deformable cells. Subcellular details are not considered.

In \cite{Mueller_PRL_2019} the propulsion of each cell is related to its deformation. For each phase field variable $\phi_i$ a Q-tensor (symmetric and trace-free) is defined by
\begin{equation*} 
		\mathbf{S}_i \!=\! \int \!\begin{bmatrix}
		\frac{1}{2} \left((\partial_y \phi_i)^2-(\partial_x \phi_i)^2\right) & \!-(\partial_x \phi_i) (\partial_y \phi_i) \\
		-(\partial_x \phi_i) (\partial_y \phi_i) & \!\!\frac{1}{2} \left((\partial_x \phi_i)^2-(\partial_y \phi_i)^2\right)
		\end{bmatrix}d \mathbf{x}
\end{equation*}
from which a continuous Q-tensor field $\mathbf{S} = \sum_{i=1}^N \mathbf{S}_i \tilde{\phi}_i$ can be constructed by interpolation, with $\tilde{\phi}_i$ a rescaled phase field variable with values in $[0,1]$. The active contribution is defined as $\mathbf{v}_i^{elo} =\int \tilde{\phi}_i \nabla \cdot \mathbf{S} \; d \mathbf{x}$. While the strength and the direction is constant within a cell, both differ between cells. Also in this approach subcellular details are not considered. However, the coupling with neighboring cells becomes stronger as they have an influence on the cell deformation.

Subcellular details, modeled as an active polar gel, have been considered in \cite{Tjhung_PNAS_2012,Ziebert_RSI_2012,Whitfield_EPJE_2014,Marth_JRSI_2015}. In these approaches for a single active droplet cell movement results form spontaneous symmetry breaking in the polarisation field of the subcellular, e.g. actin, filaments. This route to cell motility is used in \cite{Loeber_SR_2015,Marth_IF_2016} for collective cell migration and applied to simulate confluent cell structures in \cite{Wenzel_JCP_2019,Wenzel_CMAM_2021}. The free energy $\mathcal{F}$ has to be extended by a Frank-Oseen type energy
\begin{align}
    \mathcal{F}_{\mathbf{P}} = \sum_{i=1}^N \frac{1}{Pa}
    	\int_\Omega \frac{1}{2}\|\nabla\mathbf{P}_i\|^2 &+ \frac{c^{p}}{4}\|\mathbf{P}_i\|^2(-2\phi_i + \|\mathbf{P}_i\|^2) \nonumber \\ &\quad\qquad+ \beta\mathbf{P}_i\cdot\nabla\phi_i\,\text{d}\mathbf{x}
\end{align}
with polarisation field $\mathbf{P}_i$ for each cell and non-dimensional elastic parameter $Pa$. The second term ensures the unity constraint weakly in the interior, $\phi_i = 1$, and forces $\mathbf{P}_i = 0$ in the exterior, $\phi_i = -1$, with $c^p > 0$, and the third term sets an anchoring condition at the cell boundary, $\nabla \phi_i \neq 0$, with $\beta > 0$. The dynamics for each $\mathbf{P}_i$ is considered as
\begin{equation}
    \partial_t \mathbf{P}_i = - \frac{\delta \mathcal{F} }{\delta \mathbf{P}_i}, \quad i = 1, \ldots, N.
\end{equation}
In contrast to previous models \cite{Loeber_SR_2015,Marth_IF_2016,Wenzel_JCP_2019} self-advection in the evolution equation for $\mathbf{P}_i$ is omitted, to allow for better comparability with the other models. The coupling with eq. (\ref{eq:phi}) follows by defining $\mathbf{v}_i^{pol} = \mathbf{P}_i$. While the local strength remains constant, the active force is no longer equally distributed over the cell as the direction results from the subcellular polarisation field, which is strongly influenced by the geometry of the cell. On the single cell level cell movement results in the considered setting from contractile stress, see \cite{Tjhung_PNAS_2012,Marth_JRSI_2015} for details and possible modifications to generate motion by extensile stress.

\begin{figure}[h]
    \centering
    \includegraphics[width=.48\textwidth]{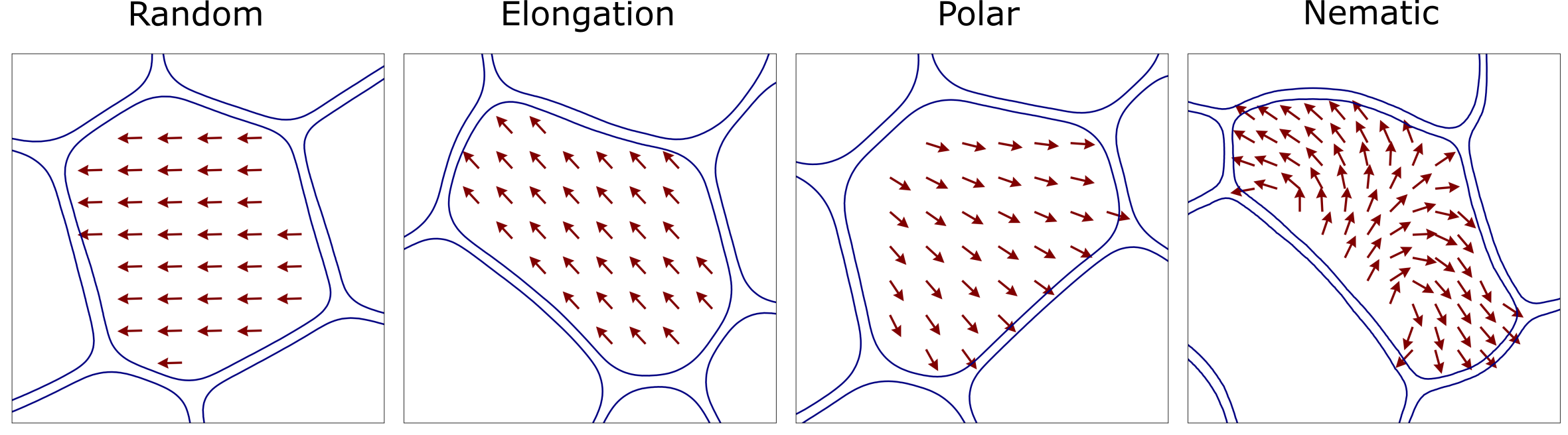}
    \caption{Visualization of representative $\mathbf{v}_i$ within one cell in confluent cell structure. From left to right $\mathbf{v}_i^{ran}, \mathbf{v}_i^{elo}, \mathbf{v}_i^{pol}$ and $\mathbf{v}_i^{nem}$. The length of the arrows is rescaled to be comparable, the blue lines indicate the $\phi_i = 0$ level sets of the considered cell and its neighbors.}
    \label{fig:schematic}
\end{figure}

As an alternative modeling approach with subcellular details we consider instead of a polar structure nematic ordering within each cell. A nematodynamic approach has been considered in \cite{Giomi_PRL_2014,Gao_PRL_2017} to model movement of a single active nematic droplet. We here simplify this approach and extend it to multiple cells. The free energy $\mathcal{F}$ is extended by a Landau-de Gennes type energy
\begin{align}
	\mathcal{F}_{\mathbf{Q}} = \sum_{i=1}^N \frac{1}{Ne}
    	\int_\Omega \frac{1}{2}\|\nabla\mathbf{Q}_i\|^2 &+
    	\text{ tr} \mathbf{Q}_i^2(-\frac{c^{n}}{2}\phi + \frac{c^{n}}{4} 
    	\text{ tr} \mathbf{Q}_i^2) \nonumber \\ &\qquad + 
    	\gamma \nabla \phi_i \mathbf{Q}_i \nabla\phi_i\,\text{d}\mathbf{x}
\end{align}
with Q-tensor field $\mathbf{Q}_i$ for each cell and non-dimensional elastic parameter $Ne$. The second term enforces $\mathbf{Q}_i = 0$ in the exterior of the cell, with $c^n > 0$ and the third term again sets an anchoring condition at the cell boundary, with $\gamma > 0$. The dynamics for each $\mathbf{Q}_i$ reads
\begin{equation}
    \partial_t \mathbf{Q}_i = - \frac{\delta \mathcal{F} }{\delta \mathbf{Q}_i}, \quad i = 1, \ldots, N,
\end{equation}
and $\mathbf{v}_i^{nem} = \nabla \cdot \mathbf{Q}_i$. In this approach the strength and direction of the active force for each cell result from subcellular structures and varies within the cell and between cells. The influence of the cell shape on the resulting movement is much stronger and a clear distinction between contractile and extensile behaviour on a single cell level not generally possible.

The differences of the models are visualized in Figure \ref{fig:schematic} for one representative cell. $\mathbf{v}_i^{ran}$ is constant within the interior of the cell. While this is also true for $\mathbf{v}_i^{elo}$, here the direction is aligned with the long axis of the cell. $\mathbf{v}_i^{pol}$ shows the typical splay instability resulting from contractile stress \cite{Tjhung_PNAS_2012,Marth_JRSI_2015} with a prefered mean orientation but otherwise constant local strength. $\mathbf{v}_i^{nem}$ shows a more complex behaviour. The underlying  instability in this model results in a rearrangement of the topological defects in the $Q$-tensor field $\mathbf{Q}_i$, which strongly depends on the geometry of the cell and leads to no prefered mean orientation of $\mathbf{v}_i^{nem}$. To explore the influence of these differences on macroscopic observables is the target of this paper.

\section{Numerics and parameter setting}

We employ a parallel and adaptive finite element method to solve the coupled system of partial differential equations for $\phi_i$ and $\mathbf{P}_i$ or $\mathbf{Q}_i$, for $i = 1, \ldots, N$, numerically. The algorithm is implemented in AMDiS \cite{Vey_CVS_2007,Witkowski_ACM_2015} and the algorithmic concepts to achieve parallel scaling with the number of cells $N$ are described in \cite{Praetorius_NIC_2017}. Briefly, they consider one core for the evolution of each cell and parallel concepts from particle methods to reduce the communication overhead due to cell-cell interaction.

We consider a constant number of cells (no cell divisions and apoptosis). First, $N = 100$ cells in a rectangular domain, $\Omega = [0,100] \times [0,100]$, with periodic boundary conditions are considered. This approximates a large confluent monolayer with no need for confinement. Second, confinements are realized in the same computational domain $\Omega$ by an implicit description using a phase field variable $\phi_\delta(\mathbf{x}) = \tanh \left((|| \mathbf{x} - \mathbf{c} || - 50 ) / (\sqrt{2}\epsilon) \right)$, where $\mathbf{c} = (50,50)^T$ and the choice of the vector norm determines the confinement shape. In particular we use 
$||\cdot||_{2}$ for a circular confinement with diameter $100$. The repulsive force of the confinement is introduced using an interaction potential as in eq. \eqref{eq:interaction} and an additional energy contribution
\begin{equation*}
    \mathcal{F}_{CO} = \sum_{i=1}^N \frac{1}{Co}\int_\Omega B(\phi_i)  w(d_\delta)\,\text{d}\mathbf{x},
\end{equation*}
following the approach of eq. \eqref{eq:IntEnergy}. Within 
the circular confinement there are $N = 106$ cells, resulting from regular initial arrangements. Considering the zero-level set of $\phi_i$ as cell boundary the resulting packing fraction is around $90\%$. The model parameters are chosen as $\epsilon = 0.15$, $In = 0.025$, $Pa = Ne = 1$, $Co = 0.004$ and $c^{p} = c^{n} = 1$. We further consider $D_r=0.1$, $\beta = 0.01$ and $\gamma = 0.1$. This allows to only vary $Ca$ and $v_0$. Other numerical parameters, such as grid resolution and time step are considered as large as possible to ensure stable behaviour and resolution of the essential physics. The grid spacing within the diffuse interface is $h \approx 0.2 \epsilon$, in the interior of each cell $h \leq \epsilon$ and in the exterior $h \leq 10\epsilon$ with increasing values for regions far away from the interior. The time step is chosen as $\tau = 0.1$.

\section{Solid-to-Liquid Transition}

We first compare collective solid-to-liquid transitions in these models. Such transitions have been observed in embryonic development and cancer progression, and may be associated with epithelial-to-mesenchymal transition in these tissues. Solid-to-liquid transitions have been extensively studied in vertex and voronoi models, see, e.g., \cite{Bi_PRX_2016}, and identified to depend on the strength of activity and cell deformability. While the deformability is typically described in these models using a shape index, we here follow \cite{Loewe_PRL_2020} and directly consider the surface tension, respectively the capillary number $Ca$. 

\begin{figure}[h]
    \centering
    \includegraphics[width=0.48\textwidth]{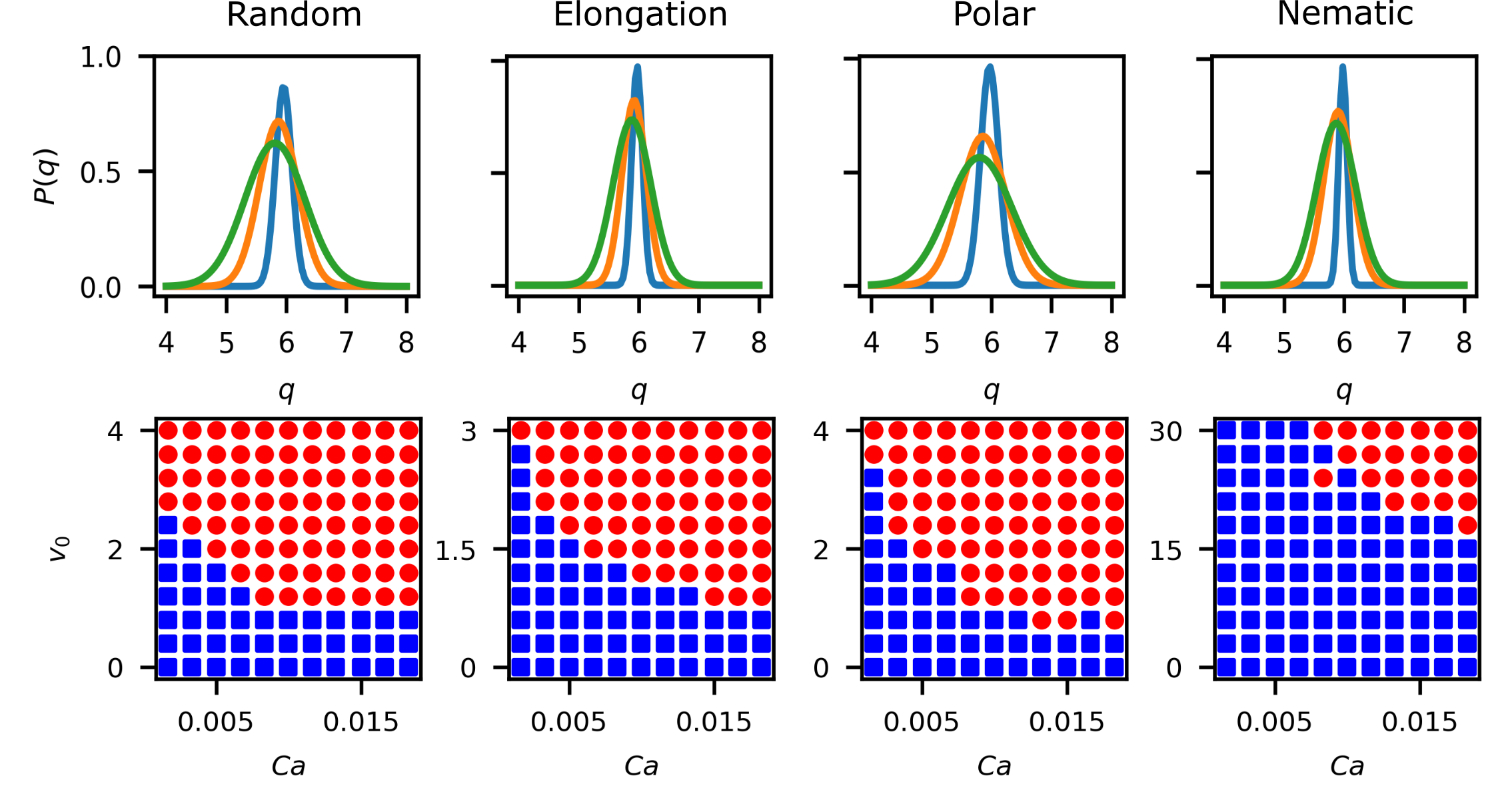}
    \caption{(first row) Coordination number probability for $Ca = 0.0148$ with low (blue), medium (orange) and high (green) values of $v_0$. For actual values see Table \ref{tab:lowmedhighActivity}. (second row) Phase diagram showing transition between solid (blue) and liquid (red) state as function of the deformability parameter (capillary number)  $Ca$ and the activity (self-propulsion strength) $v_0$. }
    \label{fig:sl1}
\end{figure}

\begin{figure}[htb]
    \centering
    \includegraphics[width=0.48\textwidth]{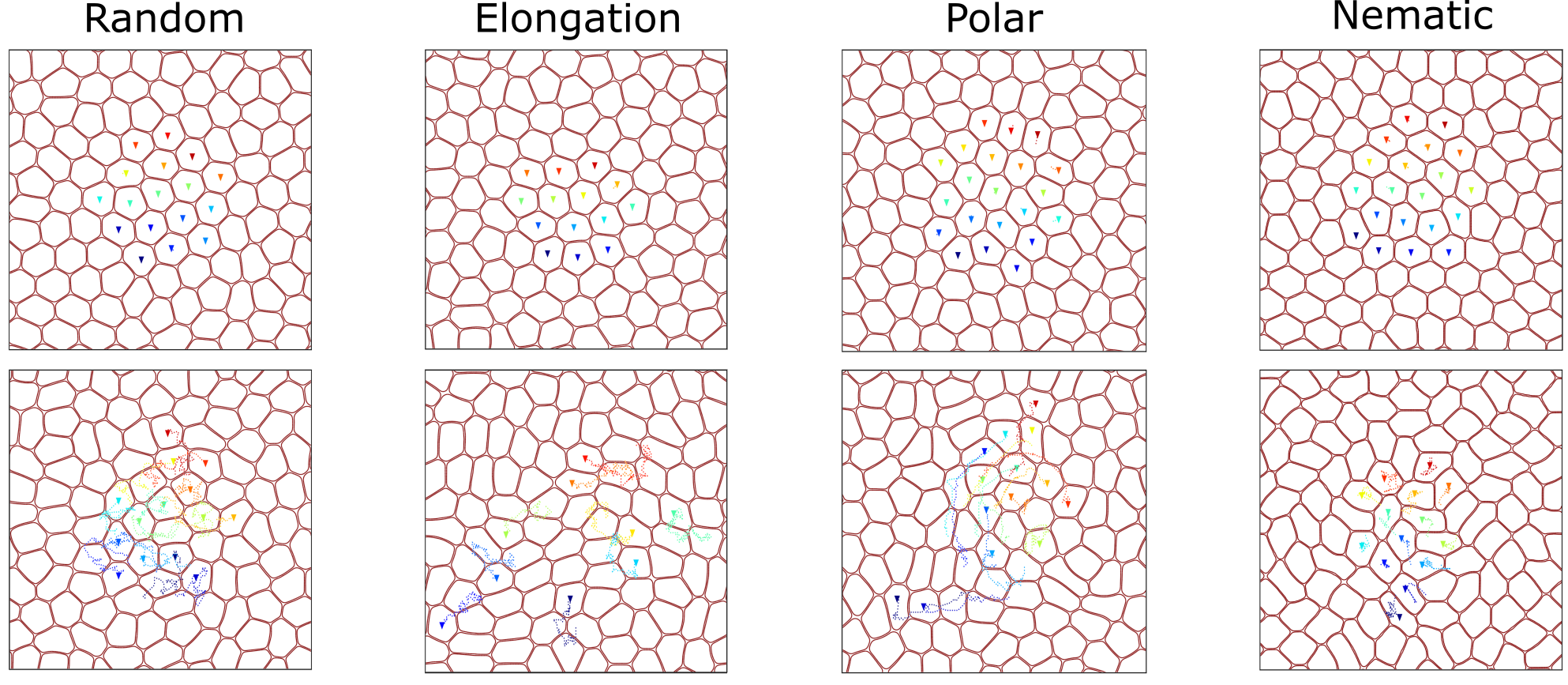}
    \caption{Snapshots of tissue morphology for liquid phase (first row) and solid phase (second row), for the four models, random, elongation, polar and nematic (from left to right). Shown are the $0$-level sets of $\phi_i$ together with cell trajectories for some time span of the cells in the center, indicating diffusion in the liquid phase and dynamical arrest due to caging in the solid phase.}
    \label{fig:sl2}
\end{figure}

To quantify the transition an easily accessible structural property, the coordination number $q$, i.e. the number of neighboring cells, is considered. Figure \ref{fig:sl1}(first row) shows the averaged distribution over all cells and all time steps for fixed $Ca$ and various $v_0$ for the four models. We consider three representative levels of activity, termed low, medium and high. The precise choice of $v_0$ depends on the mechanism of activity but is kept constant throughout the following chapters and can be found in Table \ref{tab:lowmedhighActivity}. For all models the mean value of the coordination number probability is close to $6$ and only slightly decreases with increasing activity.
\begin{table}[h]
    \begin{tabular}{c|p{1.5cm}|p{1.5cm}|p{1.5cm}|p{1.5cm}}
                 & random & elongation & polar & nematic \\ \hline
     low $\:$ & 1.2 & 1.0    & 1.2 & 21.0   \\
     medium $\:$ & 2.4 & 2.0     & 2.4 & 24.0   \\
     high $\:$ & 3.6 & 3.0     & 3.6 & 30.0   \\
    \end{tabular}
    \caption{Chosen values for $v_0$ classified as low, medium and high activity.}
    \label{tab:lowmedhighActivity}
\end{table}
The coordination number is used to identify solid-to-liquid transitions. Considering the coordination number $q$ to deviate from the hexagonal ordering, which can be expressed by the statistic variance $\mu = \sum_{q} (q-6)^2 P(q) > \theta_{P(q)}$, with $P(q)$ the discrete probability distribution obtained from counting the presence of each value $q$ and $\theta_{P(q)} = 0.001$, we can identify the solid and liquid phase. Figure \ref{fig:sl1}(second row) shows the phase diagram for the four models. Blue are regions where the observable indicates solid-like behaviour and red are regions where it indicates liquid-like behaviour. 

Although the qualitative behavior of the phase diagram is quite similar for all four models and the previous studies using vertex and voronoi models \cite{Bi_PRX_2016}, the actual quantitative results in terms of the parameter range for $v_0$ differ strongly. Both the random and the polar model are driven by a normalized vector field with a clearly preferred direction which results in a quantitatively similar behaviour. For the elongation-based and the nematic model the driving force is computed as divergence of a tensor field and thus not normalized, indicating why they have a parameter range which is different from the other two models. 

The snapshots in Figure \ref{fig:sl2} show typical cell shapes for the four models in the liquid and solid phase, respectively. The cell shapes are isotropic on average in the solid phase and anisotropic in the liquid phase, leading to differences in the number of neighbors. Also the cell tracks significantly differ, they show dynamical arrest due to caging in the solid phase and diffusion in the liquid phase. These tracks are obtained by considering the center of mass of each cell in each time step. While the solid phase is more or less identical in all four models, the liquid phase differs significantly. We will quantify these differences below.

\begin{figure}[htb]
    \centering
    \includegraphics[width=.48\textwidth]{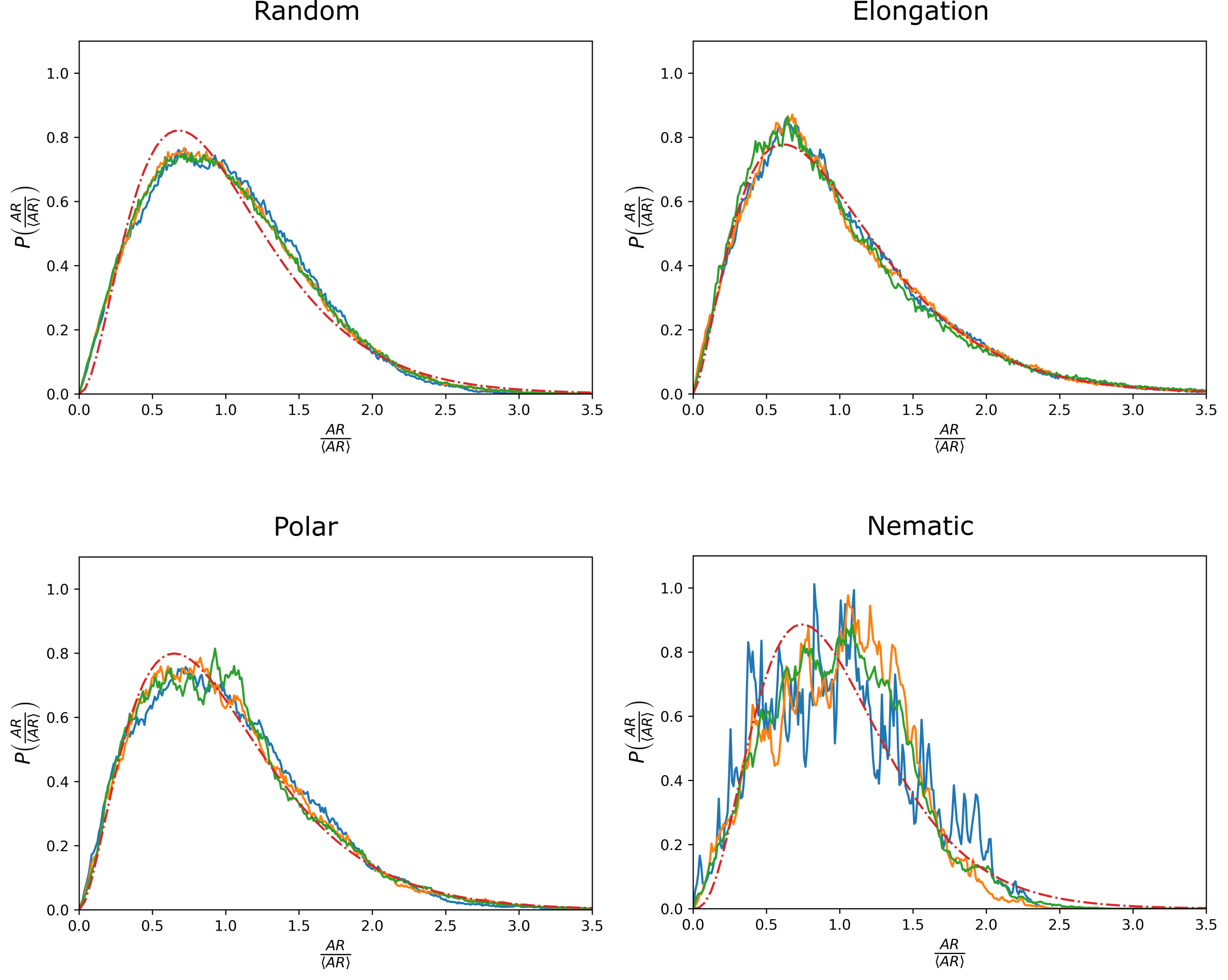}
    \caption{Shape variability for the four models using the rescaled parameter $x = \frac{AR}{\langle AR \rangle}$ for different points of the phase diagrams in Figure \ref{fig:sl1} with low (blue), medium (orange) and high (green) activity. Fitted values for the PDF (red) are computed using data for all three values $v_0$ resulting in $k^{ran}=3.11$ ,$k^{elo}=2.59$, $k^{pol}=2.84$ and $k^{nem}=3.88$.}
    \label{fig:shape}
\end{figure}

Following \cite{Atia_NP_2018}, it should also be possible to identify common generic features over a wide range of the phase diagrams. Considering the shape variability of the cells, which is expressed by the aspect ratio $AR$ of the long and the short cell-axis, the largest and the smallest eigenvalue of $\mathbf{S}_i$, and rescaling $x = \frac{AR}{\langle AR \rangle}$, with $\langle \cdot \rangle$ the average value, leads to an empirically proposed universal k-Gamma distribution \cite{Atia_NP_2018} with probability distribution function $PDF(x,k) = \frac{k^k}{\Gamma(k)} x^{k-1} e^{-kx}$ with Legendre Gamma-function $\Gamma(k)$. This distribution is fully described by the parameter, $k$, and has a mean of unity. Across diverse epithelia systems, including Madin-Darby canine kidney (MDCK) cells, Human broncial epithelia cells (HBECs) and the Drosophila embryo during ventral furrow formation, \cite{Atia_NP_2018} shows that this equation pertain with $k$ in a narrow range between $2$ and $3$, which indicates universality. This result provides the first quantitative comparison for the different phase field models. Figure \ref{fig:shape} shows the distribution for all four models together with maximum-likelihood-estimation fits for $PDF(x,k)$ for three different parameters $v_0$ and $Ca = 0.018$. The corresponding $k$-values for each $v_0$ are shown in Table \ref{tab:parameterK}.

\begin{table}[h]
\begin{tabular}{c|p{1.5cm}|p{1.5cm}|p{1.5cm}|p{1.5cm}}
 $v_0$             & random & elongation & polar & nematic \\ \hline
 low $\:$ & 3.18 & 2.68    & 2.82 & 3.23   \\
 medium $\:$ & 3.12 & 2.60     & 2.87 & 4.20   \\
 high $\:$ & 3.04 & 2.49     & 2.85 & 4.21   \\
\end{tabular}
\caption{k-Gamma parameter fit for different models. (low, medium, high) corresponds to the values used in Figure \ref{fig:shape}. The $k$-values are obtained with maximum-likelihood-estimation fits for $PDF(x,k)$ for one simulation run over the whole time.}
\label{tab:parameterK}
\end{table}

Indeed, in accordance with the experimental results in \cite{Atia_NP_2018}, the data can be described by a k-Gamma distribution and $k$ does not vary strongly within each model for the considered parameters. However, the $k$-values differ between the four models. While both the elongation-based and the polar model have values within the experimentally predicted universal range between 2 and 3, both the random and the nematic model are slightly above with the latter one leading to the largest values. These larger values for $k$ in the nematic model become evident from the construction of the model, where active forces enhance elongation which is also apparent in Figure \ref{fig:schematic}. This also explains why $k$ is growing for larger values of $v_0$ only in the nematic model while it stays approximately constant in all others. The larger fluctuations in the polar and especially the nematic model can be explained by the stronger coupling between shape changes and active forces.

These differences in the shape variability of the cells provide a first indication on the dependency of macroscopic observables on the microscopic details considered in each model.

\section{Liquid Phase}

We now only concentrate on the liquid phase and compare the four models with other statistical observables of experiments. We consider vorticity correlation functions and statistical data on topological measures, such as number of neighboring cells and distributions of topological defects. In order to be comparable we parameterize all models to fulfill one common topological measure. While the deformability parameter, the capillary number, can be chosen as $Ca=0.018$ in all models, the activity parameter $v_0$ differs to model the same physical state in the phase diagram. We consider the variance in the coordination number $\mu = 0.4$ as reference value. This value corresponds to a value measured in Drosophila embryos \cite{Blankenship_DevCell_2006} at an early stage of development (up to stage 7 before intercallation). Using $v_0^{ran} = 1.3, v_0^{elo} = 0.6, v_0^{pol} = 1.3$ and $v_0^{nem} = 19.0$, we obtain $\mu = 0.41, 0.40, 0.44$ and $0.45$, respectively. The resulting configurations with these parameters are considered as comparable physical states.

\subsection{Rosette formation}

Multicellular rosettes or higher-order vertices, where four or more cells meet, have been found in many tissues \cite{Harding_Dev_2014}. The importance of cellular rosettes has been widely recognized and they have been proposed as an efficient mechanism for tissue remodeling. In \cite{Yan_PRX_2019} the influence of
rosettes on the mechanics of a confluent tissue is studied using a generalized vertex model. While in these models the formation of rosettes requires an adhoc collapse of cell edges, T1 junctions and rosettes form naturally within our multiphase field models \cite{Wenzel_JCP_2019}. The rosette ratio, the fraction of all vertices that connect more than three cells, is shown in Table \ref{tab:ratio}. 

\begin{table}[h]
\begin{tabular}{c|p{1.5cm}|p{1.5cm}|p{1.5cm}|p{1.5cm}}
              & random & elongation & polar & nematic \\ \hline
rosette ratio $\:$ & 4.8\% & 2.6\%     & 5.5\% & 1.3\%   \\
\end{tabular}
\caption{Rosette ratio for different models, considered for one simulation over the whole time for comparable physical states.}
\label{tab:ratio}
\end{table}

Experimental data for the rosette ratio for the corresponding early stage of development (up to stage 7 before intercallation) in Drosophila embryos \cite{Blankenship_DevCell_2006} show values between $5 \%$ and  $6 \%$, which is reproduced by the polar model. The random model leads to a ratio which is only slightly below. The other two models lead to significantly lower values. In later stages of development this ratio is drastically increased. For the Drosophila embryos the peak fraction of cells in rosettes at a single time point is $61\%$ \cite{Blankenship_DevCell_2006}. But this corresponds to a different physical setting, for which the models have not been calibrated. Other data, e.g., in Zebrafish embryo \cite{Hava_JCS_2009} report a ratio of $1.8 \%$. However, also for these data the models are not calibrated. Even if only one physical state is considered, which allows for a calibrated comparison with experimental data, the results strongly differ between the four models. These differences on the rosette ratio indicate a further dependency on the mechanism of propulsion. 

\subsection{Nematic order and topological defects}

\begin{figure}[htb]
    \centering
    \includegraphics[width=.48\textwidth]{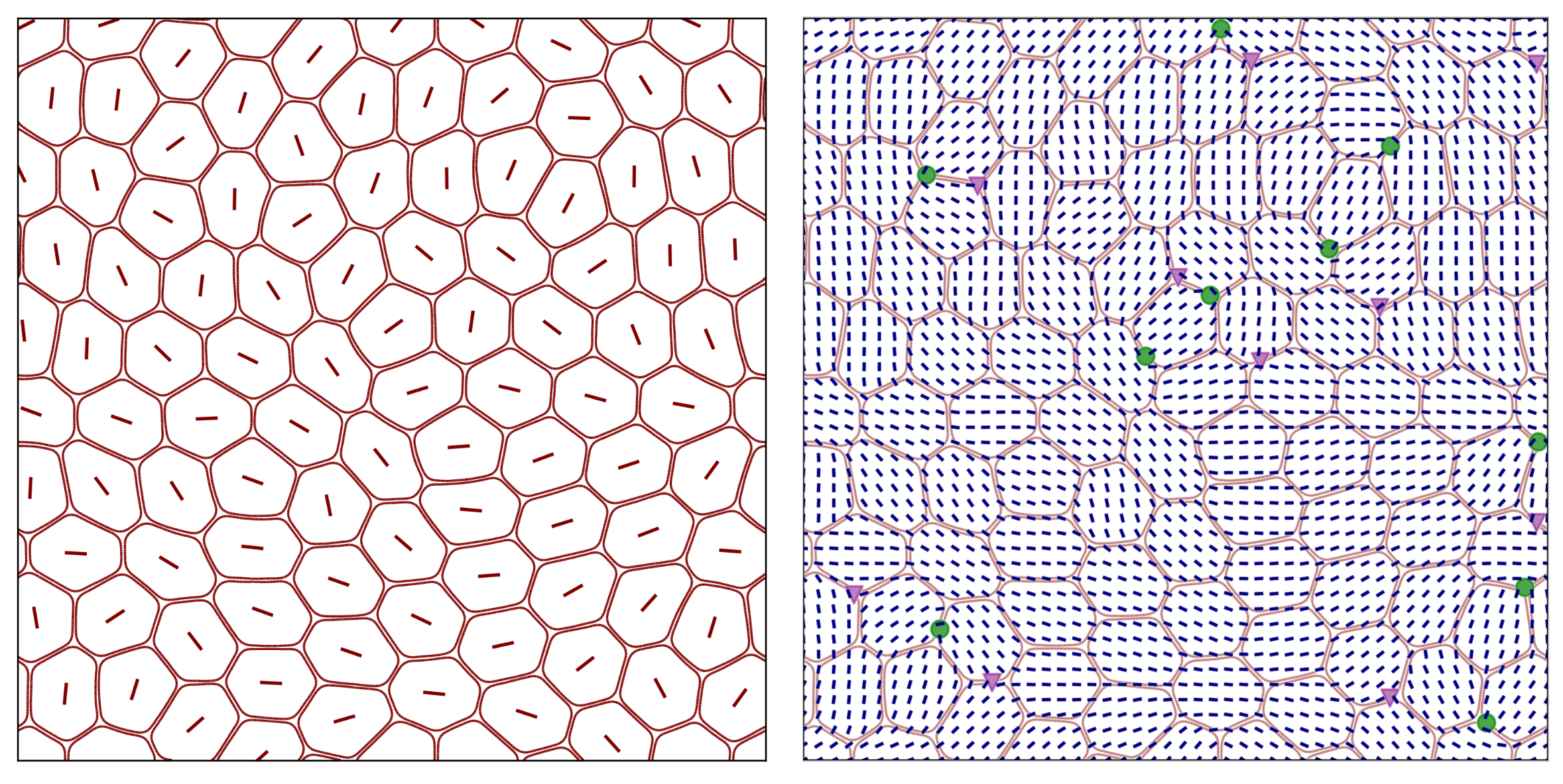}
    \caption{(left) Tissue morphology represented by $\phi_i = 0$ level lines, together with normalized largest eigenvector of $\mathbf{S}_i$ corresponding to orientation of long cell-axis in center of mass of each cell (red lines). (right) Global nematic field $\mathbf{S}$ obtained by interpolation of $\mathbf{S}_i$, represented by director field (blue lines) and $+\frac{1}{2}$ (green) and $-\frac{1}{2}$ (purple) defects. The $\phi_i = 0$ level lines are shown to indicate the position of defects in relation to the morphology.}
    \label{fig:defect}
\end{figure}

Many cellular systems in its liquid phase display properties of active liquid crystals, such as local nematic alignment and the appearance of topological defects. For MDCK cells it has been shown that these defects can control death and extrusion in cell monolayers \cite{Saw_N_2017}. This relation has been used to model collective cell migration on a multicellular scale, see \cite{Doostmohammadi_NC_2018}. In \cite{Mueller_PRL_2019} the formation of nematic order is addressed using a multiphase field model. We follow this procedure and compare the emergence of global nematic order and the proliferation of topological defects in the orientation field in all four models. We thereby determine nematic order from cell deformations and compute the local Q-tensor $\mathbf{S}_i$ for each phase field $\phi_i$. The eigenvalues and eigenvectors of $\mathbf{S}_i$ measure the strength and orientation of the main deformation axis of cell $i$. Interpolating $\mathbf{S}_i$, as described for the elongation-based model above, defines a global Q-tensor $\mathbf{S}$. Different methods exist to identify topological defects in $\mathbf{S}$, they have been compared in \cite{Wenzel_CMAM_2021}. We here consider a physics based approach, which addresses degenerated points of $\mathbf{S}$ to identify the location of defects and the sign of $\delta = \partial_x S_{11} \partial_y S_{12} - \partial_y S_{11} \partial_x S_{12}$ to distinguish between $+\frac{1}{2}$ and $-\frac{1}{2}$ defects. Figure \ref{fig:defect} illustrates the process for one snapshot. 

\begin{figure}[htb]
    \centering
    \includegraphics[width=.48\textwidth]{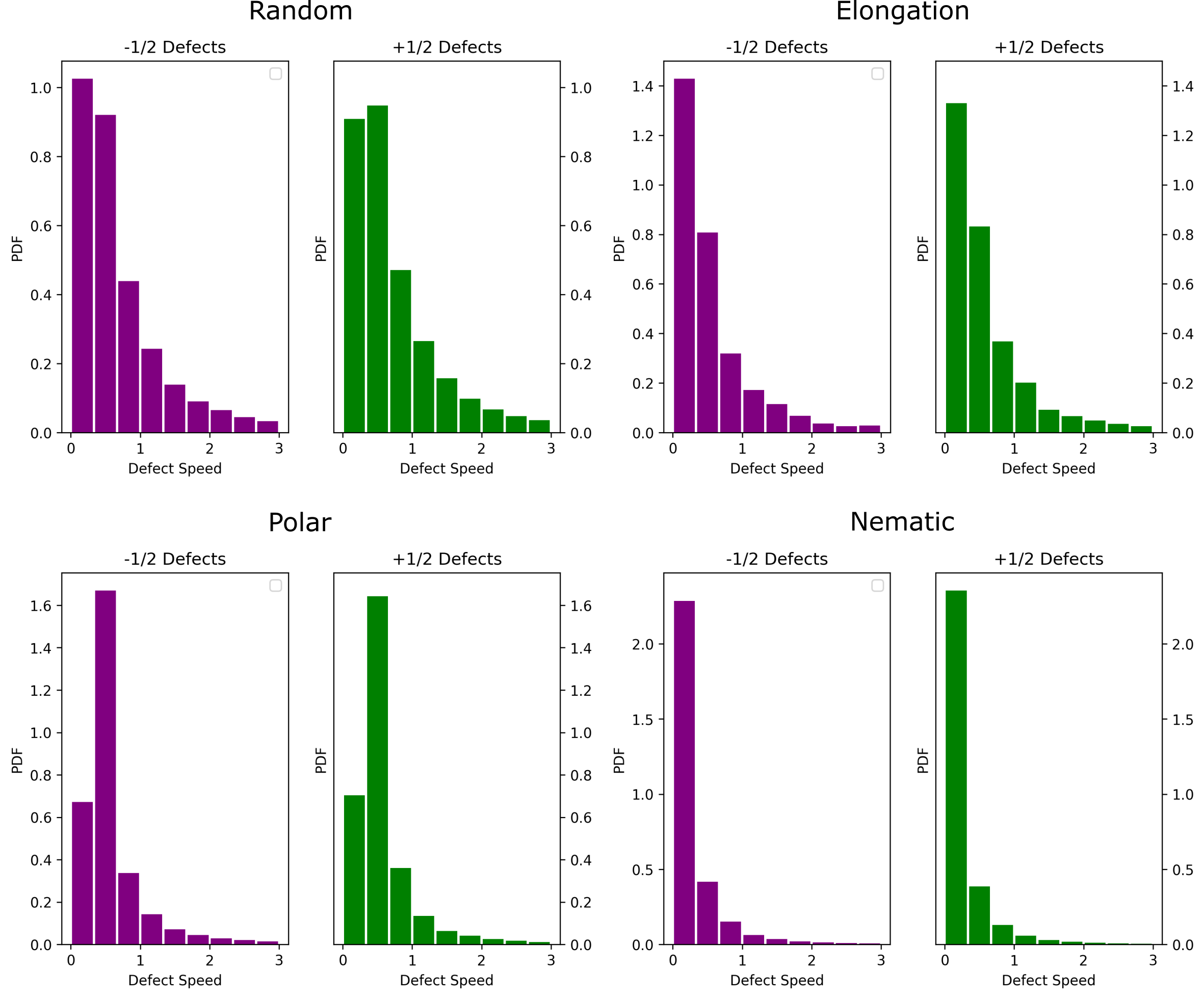}
    \caption{Velocity distribution of topological defects ($+ \frac{1}{2}$ and $- \frac{1}{2}$) for all four models: random, elongation, polar and nematic, from top-left to bottom-right.}
    \label{fig:vel}
\end{figure}

Localisation and identification of defects is done in each time step. In order to connect the defects from frame to frame we consider a particle tracking algorithm \cite{Sbalzarini_JSB_2005}, available in ImageJ/FiJi \cite{Schindelin_NM_2012} and shown to reliably consider different defect types and defect appearance and disappearance \cite{Wenzel_CMAM_2021}. This allows to statistically examine the velocity distribution of topological defects in all four models, see Figure \ref{fig:vel}. 

\begin{figure}[htb]
    \centering
    \includegraphics[width=.48\textwidth]{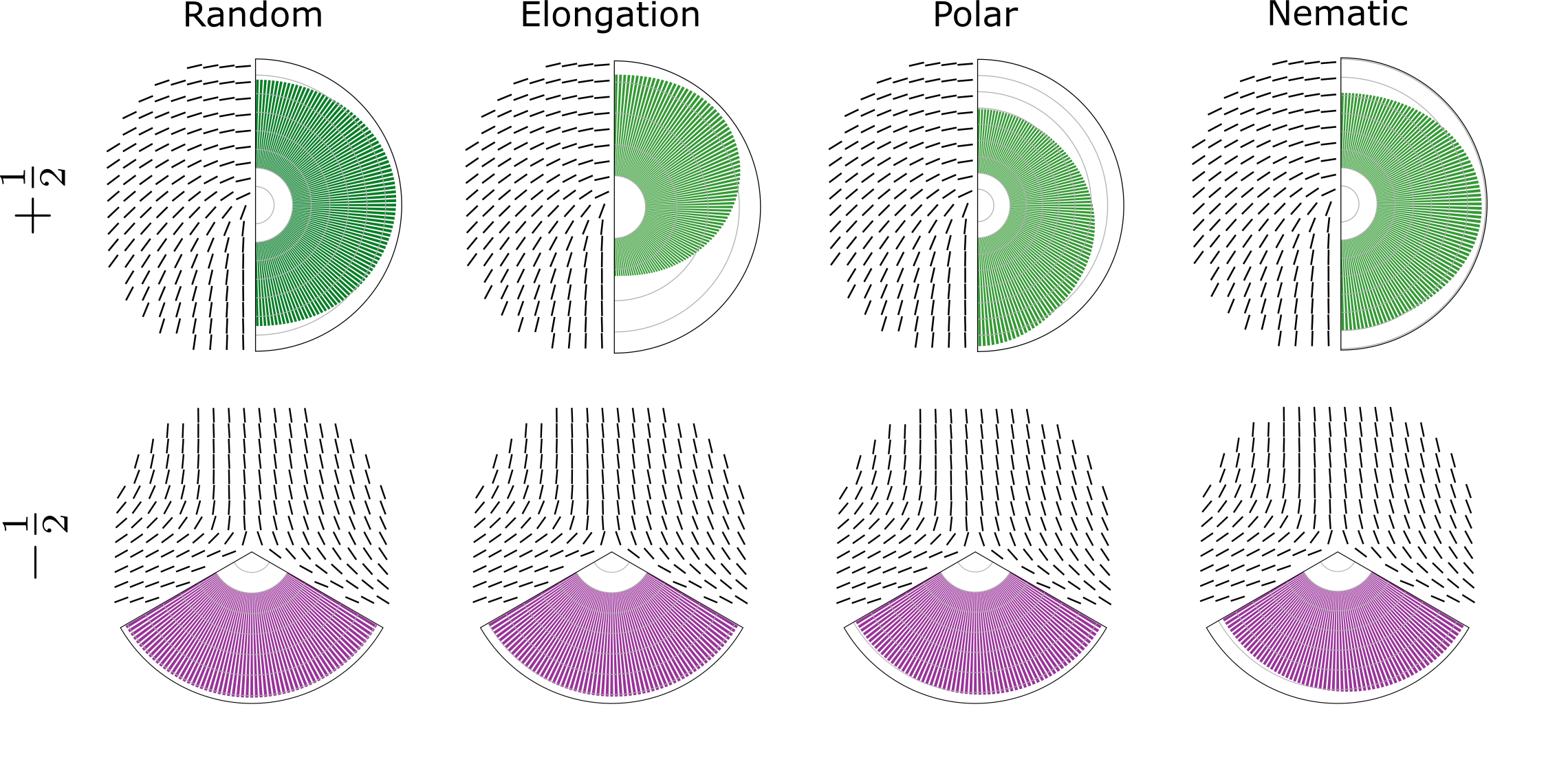}
    \caption{Distribution of direction of motion with respect to symmetry properties of $+ \frac{1}{2}$ (top) and $- \frac{1}{2}$ (bottom) defects for all four models. A schematic description of the defects defines the considered symmetry.}
    \label{fig:vel_def}
\end{figure}

While these distributions strongly differ between the four models, in all models the velocity distribution of $+ \frac{1}{2}$ and $- \frac{1}{2}$ defects is similar. This qualitative difference with coarse grained active nematodynamics and experimental data, e.g., for active microtubule networks, which indicate a difference in the velocity distribution between the different types of defects, see \cite{DeCamp_NM_2015,Oza_NJP_2016}, has already been found in \cite{Wenzel_CMAM_2021}. Detailed data on the velocity distribution of $+ \frac{1}{2}$ and $- \frac{1}{2}$ defects for epithelia cell cultures are not separately available. However, for HBECs, \cite{Blanch_PRL_2018} indicates no apparent quantitative differences between both types of defects in terms of their trajectories on long time scales, which might support the simulation results. However, differences between $+ \frac{1}{2}$ and $- \frac{1}{2}$ defects become evident if the direction of the defect velocity is correlated with the local properties of the defect. Figure \ref{fig:vel_def} shows the distribution of directions with respect to symmetry properties of $+ \frac{1}{2}$ and $- \frac{1}{2}$ defects. 


While the velocity of $- \frac{1}{2}$ defects do not show any preferred orientation for all models, which supports the passive (diffusive) role of these defects, the velocity of $+ \frac{1}{2}$ defects is strongly correlated with the head or the tail of the defect. Only the random model does not show this property. All other models support the active role of $+ \frac{1}{2}$ defects in active nematic systems \cite{Doostmohammadi_NC_2018}. The elongation model shows a strong correlation of the direction of movement with the head of the defect, indicating extensile behaviour. The polar and nematic model show a stronger correlation with the tail of the defect, indicating contractile behaviour. For a detailed discussion of these relations in active nematics we refer to \cite{Giomi_PTA_2014}. 

\begin{figure}[htb]
    \centering
    \includegraphics[width=.48\textwidth]{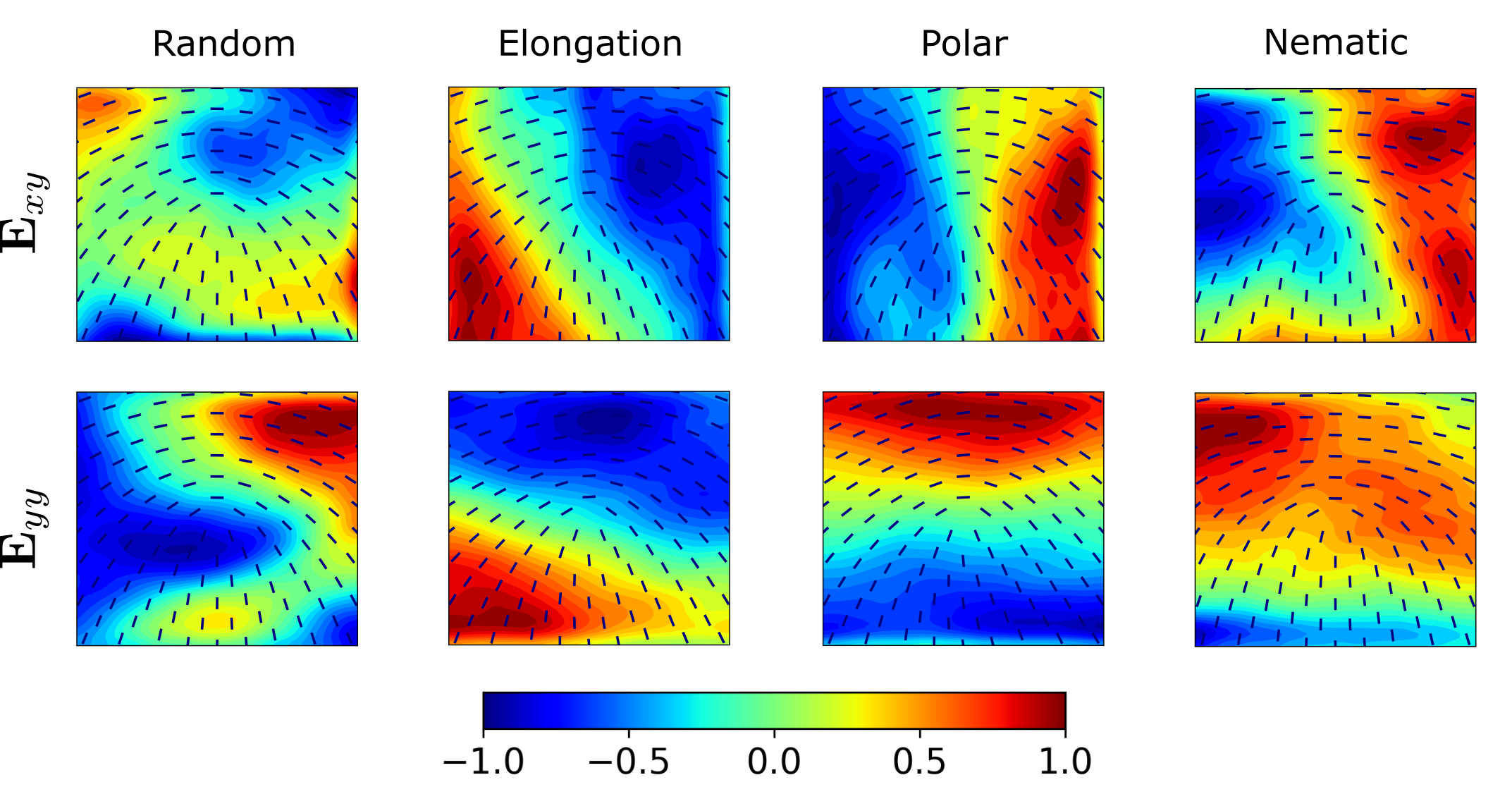}
    \caption{Average fields for both the $xy$ component (top) and the $yy$ component (bottom) of the strain rate tensor $\boldsymbol{E}$ in the vicinity of $+ \frac{1}{2}$ defects for all models: random, elongation, polar, nematic from left to right. Each plot shows a box of dimension $8\times 8$ centered at the defect core. The averaged is taken over data of more than $3000$ defects for each model.}
    \label{fig:strain}
\end{figure}

To further elaborate on the hypothesis that microscopic details on the single cell level determine the mechanical properties of the system, we compute the strain rate tensor in the vicinity of $+\frac{1}{2}$ defects. The essential quantity is the velocity $\mathbf{v}$ obtained by linear interpolation of the cell velocities $\mathbf{v}_i^{cell}$, which are computed from the movement of the center of mass of the cells. The strain rate tensor $\boldsymbol{E} = \frac{1}{2} (\nabla \mathbf{v} + (\nabla \mathbf{v})^T)$ is defined in the vicinity of the $+ \frac{1}{2}$ defects and averaged after appropriate reorientation over all defects. Figure \ref{fig:strain} shows the averaged fields for all models. 

With the exception of the random model, which does not show any significant pattern, the other models support our hypothesis. The elongation model leads to patterns characteristic for extensile systems, while the polar and nematic model show patterns characteristic for contractile systems. The strain rate along the tail-to-head direction (yy-strain) shows negative (positive) values at the head indicating the presence of compression (extensional deformation). The presence of both types is know from experiments, e.g., epithelial \cite{Saw_N_2017,Blanch_PRL_2018} or neural progentior \cite{Kawaguchi2017TopologicalDC} monolayers behave as an extensile system, while monolayers of fibroblasts \cite{Duclos2017TopologicalDI} behave as a contractile system. The extensile behavior of the elongation model has already been found in \cite{Mueller_PRL_2019}. The model is constructed to elongate the cell further along its long axis, see definition of $\mathbf{v}^{elo}$ and Figure \ref{fig:schematic}. Due to the interaction of cells this behaviour leads to extensional deformations. For the polar model the contractile stress on the single cell level also generates contractile behavior at the collective level. In the nematic model the behavior on the single cell level strongly depends on the shape of the cell. However, the collective behavior shows contractile patterns. 

The differences between epithelial and mesenchymal cells, which show extensile and contractile behaviour at the collective level, respectively, have been explored in \cite{balasubramaniamEtAl2021}. The different mechanical behavior is associated with strong cell-cell adhesion in epithelial monolayers, which allows for active intercellular force transmission. Weakening this intercellular adhesion results in contractile behavior at the collective level, consistent with the contractile stress on the single cell level. The multiphase field model used in  \cite{balasubramaniamEtAl2021} to confirm these findings combines features of our elongation, polar and nematic model. However, the mechanical behavior on the single cell level remains unclear. The purpose of our study is to first fully understand the emerging behaviour of each microscopic effect separately, before these effects are combined. In any case the experiments in \cite{balasubramaniamEtAl2021} suggest, that additional cell-cell adhesion can change the collective mechanical properties from contractile to extensile in the polar and nematic model. Also in \cite{balasubramaniamEtAl2021} the average velocity in monolayers is compared between extensile and contractile systems at similar density, with larger velocities for the extensile system. Comparing the velocity distribution in Figure \ref{fig:vel} between the elongation model (extensile) and the polar and nematic models (contractile) we find for the average velocity of $+ \frac{1}{2}$ defects a consistent behavior, see Table \ref{tab:velocity}. The even higher number for the random model results from the large velocity fluctuations in this model. 

\begin{table}[h]
\begin{tabular}{c|p{1.5cm}|p{1.5cm}|p{1.5cm}|p{1.5cm}}
              & random & elongation & polar & nematic \\ \hline
velocity $\:$ & $0.954$  &    $0.710$  & $0.653$ & $0.294$  \\
\end{tabular}
\caption{Average velocities for $+ \frac{1}{2}$ defects for all 4 models.}
\label{tab:velocity}
\end{table}

\subsection{Active turbulence and vorticity correlation}

\begin{figure}[t]
    \centering
    \includegraphics[width=.48\textwidth]{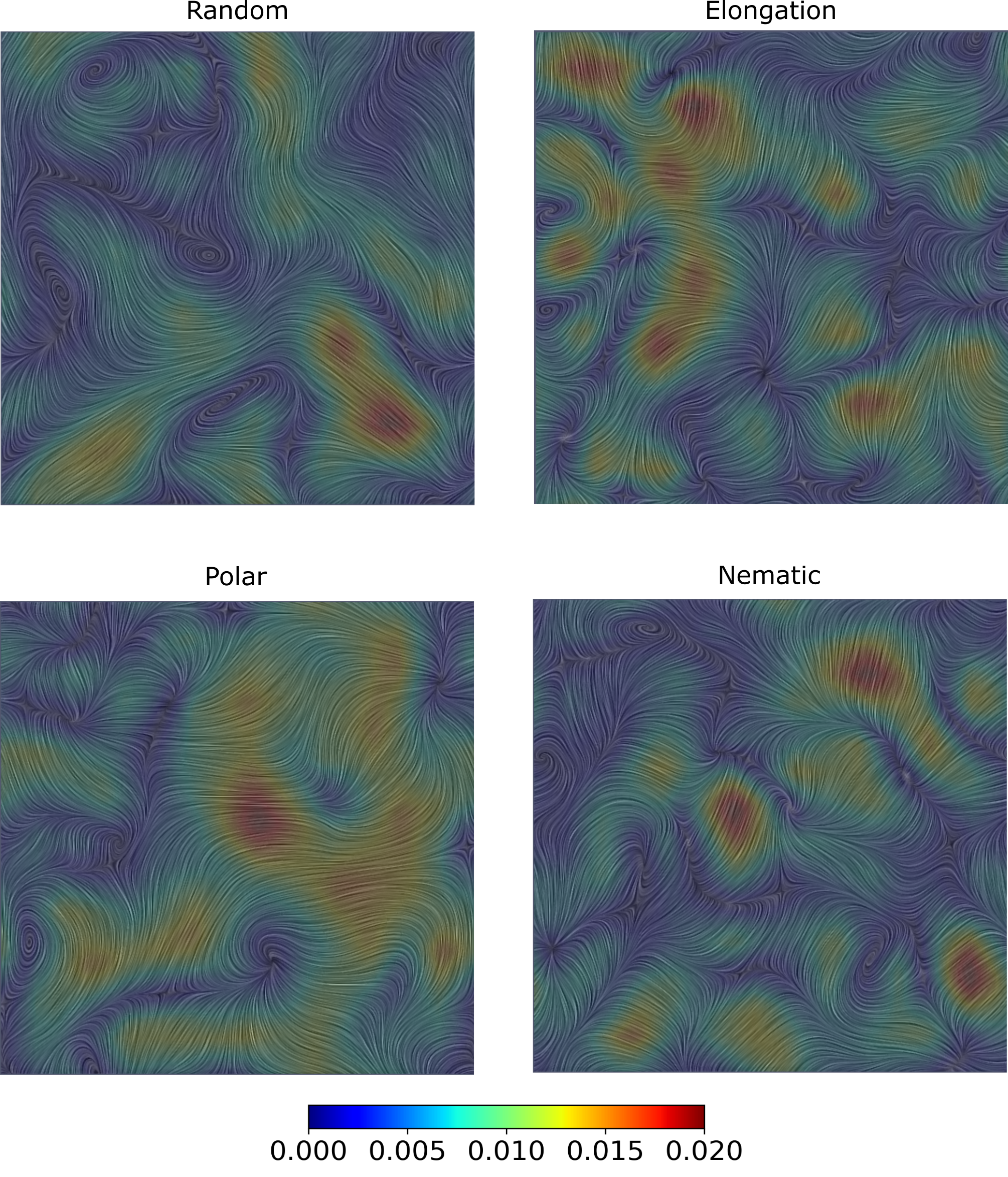}
    \caption{LIC visualisation of cell dynamics for the four models: random, elongation, polar and nematic (from left to right). Color represents the magnitude of the velocity with the same scaling for all models.}
    \label{fig:lic}
\end{figure}

\begin{figure}[h]
    \centering
    \includegraphics[width=.46\textwidth]{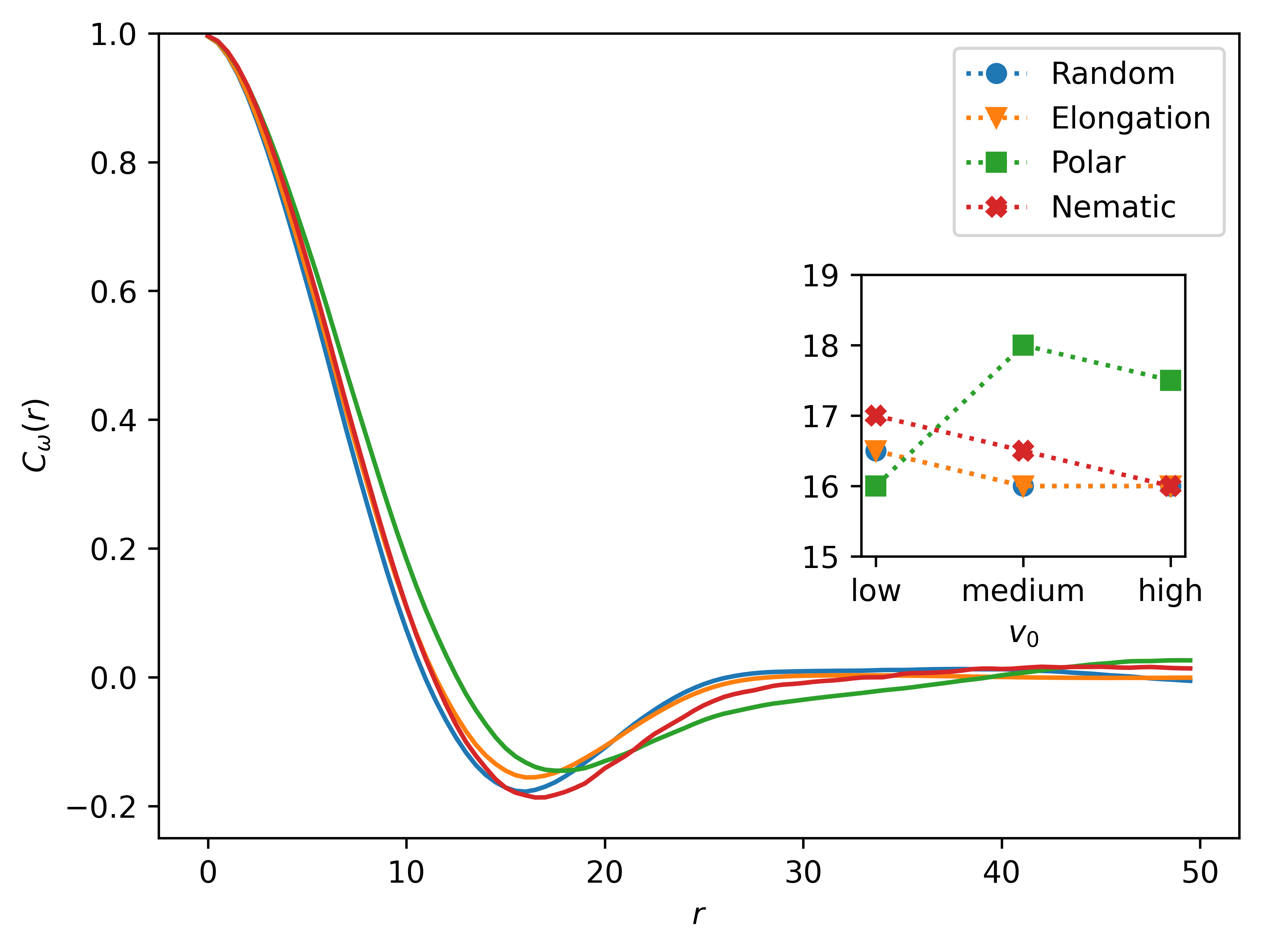}
    \caption{Vorticity-vorticity correlation function $C_\omega(r) = \langle \omega(r) \omega(0) \rangle / \langle \omega(0)^2\rangle$ depending on the distance $r$ for all models. The data is averaged over 3 simulations with the "low" values for the self-propulsion velocity, see Table \ref{tab:lowmedhighActivity}. The other values lead to qualitatively similar results. The inlet shows the position of the minima for the other activity values.}
    \label{fig:corr}
\end{figure}

For large enough activities also flow patterns reminiscent of active turbulence can be found in confluent cell structures. Examples are collectively migrating MDCK cells, fibroblastlike normal rat kidney (NRK) cells and HBKCs, which show long-range flows and patterns of vorticity, see e.g. \cite{Petitjean_BPJ_2010,Blanch_PRL_2018}. In models for active liquid crystals such turbulent states emerge as a result of spontaneous defect pair creation. In \cite{Mueller_PRL_2019} the velocity field is also analysed for a multiphase field model. We here follow this approach and compare the four models. Figure \ref{fig:lic} shows snapshots of the cell dynamics, visualized using LIC to highlight the active turbulent character of the dynamics. The vorticity is computed from the velocity field as $\omega = \mbox{curl } \mathbf{v}$. 
A vorticity-vorticity correlation function can be computed, which is shown in Figure \ref{fig:corr}. It has a well-defined minimum and thus confirms a macroscopic length scale for long range flows, which is mediated by the activity of the individual cells and their interaction. This length scale more or less coincides for the four models and only slightly changes with the strength of activity. 

\begin{figure}[h]
    \centering
    \includegraphics[width=.48\textwidth]{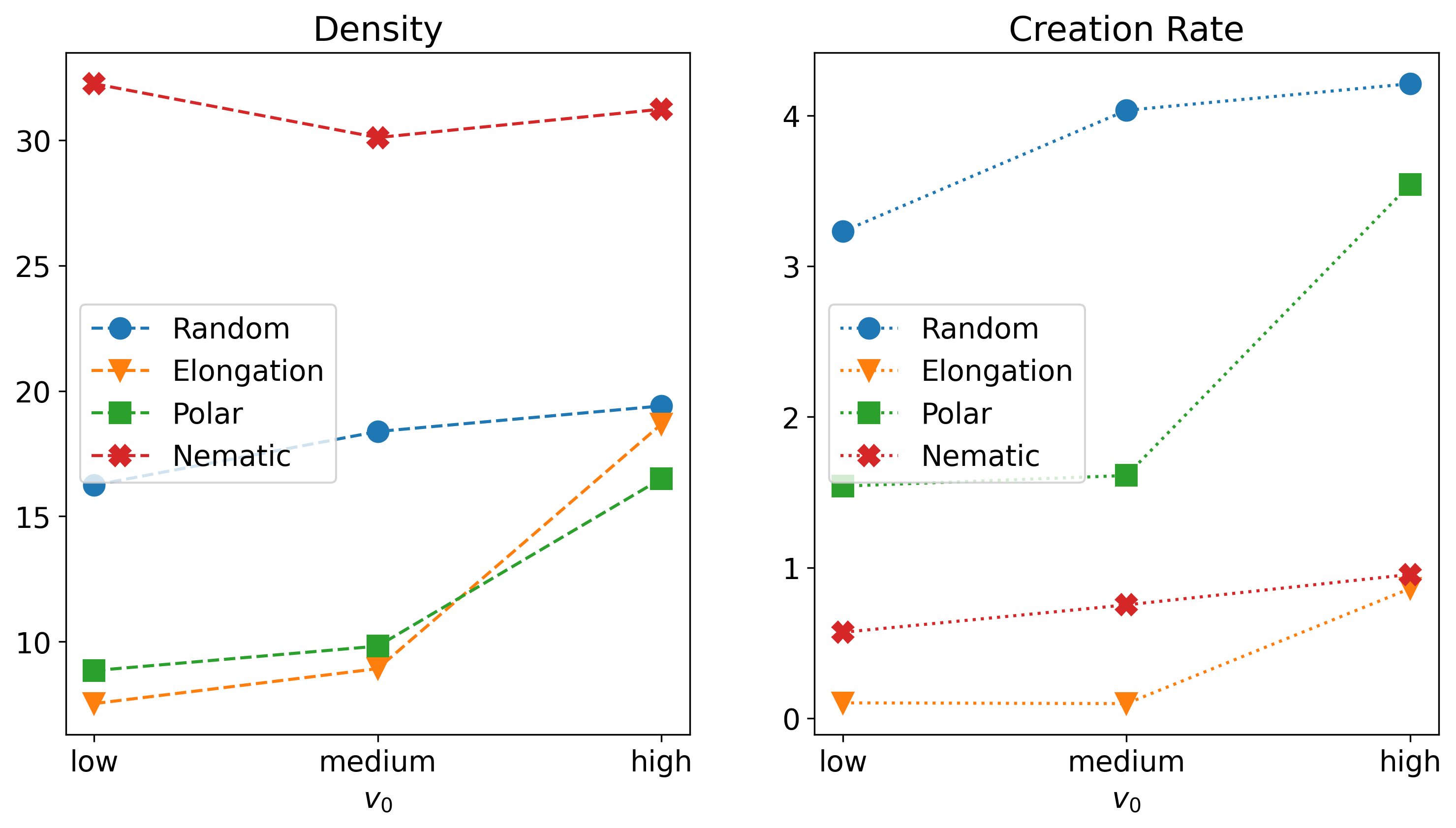}
    \caption{Averaged defect density (left) and creation rate (right) as function of  activity with low, medium and high values defined in Table \ref{tab:lowmedhighActivity}.}
    \label{fig:rate}
\end{figure}

As the turbulent collective flow is characterized by the spontaneous emergence of mesoscopic vortices and nematic defects, we also analyse the defect density and creation rate, see Figure \ref{fig:rate}. Simulations for active nematics and experiments on MDCK cells (with activity reduced by blebbistatin) \cite{Saw_N_2017}, show a linear dependency of the defect density on activity. This behaviour is qualitatively reproduced by the random, elongation and polar model. The nematic model shows slight deviations with no consistent slope. The behaviour correlates with the defect creation rate. Comparing the absolute values, the nematic model leads to significantly larger defect densities but lower creation rates, which indicates stronger persistence of defects. In contrary the random model leads to significantly larger creation rates, which might be explained by the random component of the model.    

\subsection{Confinement}

While all investigations above consider a large confluent monolayer, we are now concerned with the influence of confinement on the emerging macroscopic behaviour. The first multiphase field simulations of such situations consider persistent rotational motion of two cells \cite{Camley_PNAS_2014} on adhesive micropatterns. In this model $\mathbf{v}_i$ follows from a reaction-diffusion equation to be solved within each cell. The emerging patterns in concentration of Rho GTPase define a polarity, which determines strength and direction of motion. For more detailed modeling approach in this direction we refer to \cite{Marth_JMB_2014} and the references therein. Already these simulations, which consider the simplest possible collective motion, show a strong dependency on subcellular features on the emerging behaviour. Recent studies with more cells in a rectangular confinement could reproduce sustained oscillation experimentally observed for MDCK cells, human keratinocytes (HaCat) and enterocytes (CaCo2) \cite{Peyret_BJ_2019}. The considered multiphase field models in these studies are related to the elongation model \cite{Mueller_PRL_2019} and the polar model \cite{Wenzel_CMAM_2021}. 



\begin{figure}[t]
    \centering
    \includegraphics[width=.48\textwidth]{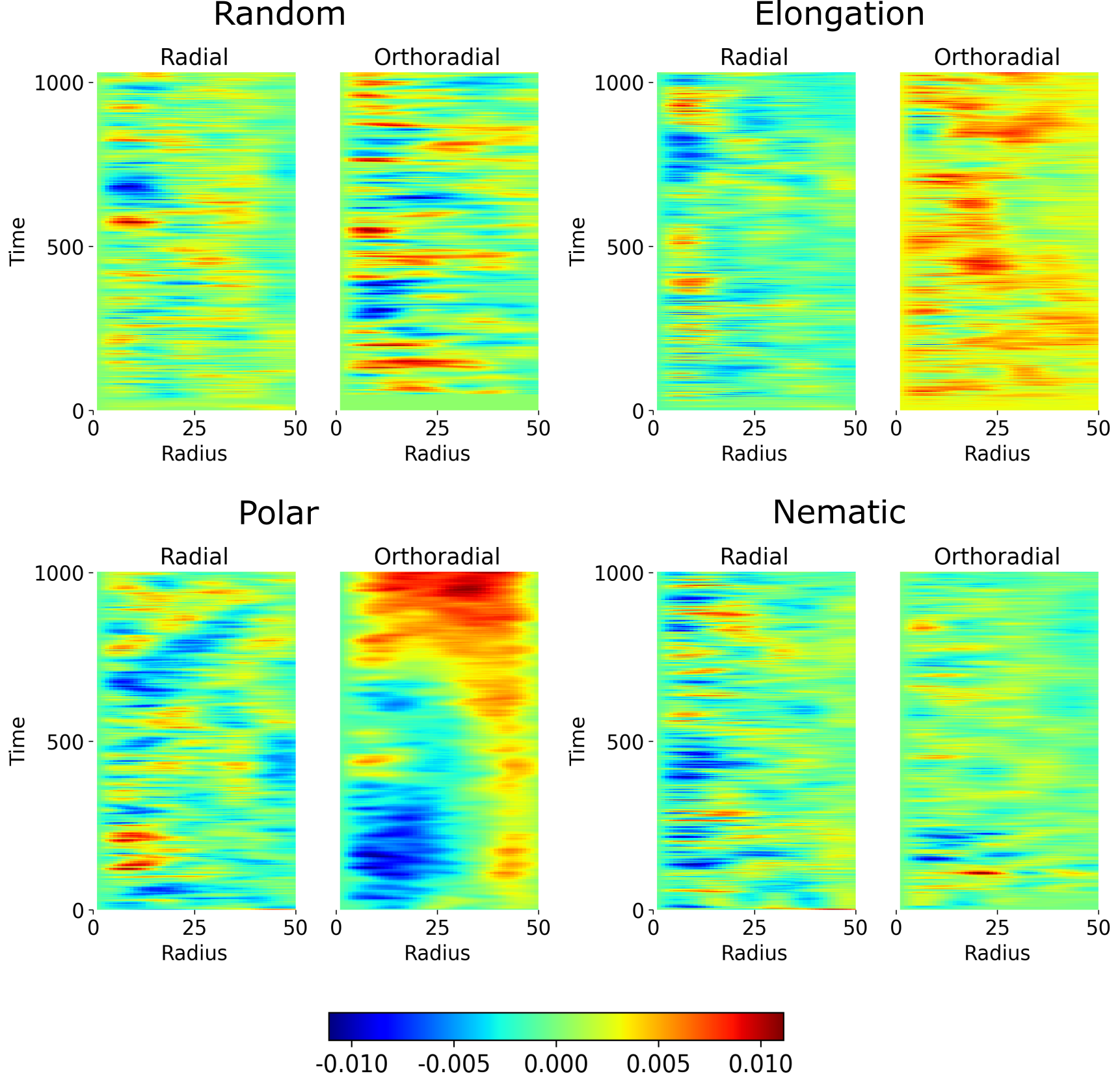}
    \caption{Kymographs of radial and orthoradial velocity components for the four models: random, elongation, polar and nematic, from left to right. Corresponding movies of the evolution are provided in the Electronic Supplement.}
    \label{fig:kymo}
\end{figure}

To compare the four models we focus on experiments for MDCK cells in circular confinements \cite{Deforet_NC_2014}. They show that confined epithelia exhibit collective low-frequency radial displacement modes and rotational motion, which was partly reproduced in corresponding particle-based simulations \cite{Hakim_RPP_2017}. The circular geometry allows to split the velocity $\mathbf{v}$ into radial and orthoradial components, which can be averaged over all angles to obtain their mean spatial distributions. These values are shown in Figure \ref{fig:kymo} for all four models. While the radial component is qualitatively similar in all four models, the orthoradial component qualitatively differs between the models. Only the polar model could reproduce the rotation of the monolayer as a whole and a change in direction of the interior part, which is assumed to be responsible to a comparable size of the confinement and the spatial scale resulting from the vorticity-vorticity correlation in Figure \ref{fig:corr}. The simulations are performed with the "medium" values in Table \ref{tab:lowmedhighActivity}. 

\begin{figure}[t]
    \centering
    \includegraphics[width=.48\textwidth]{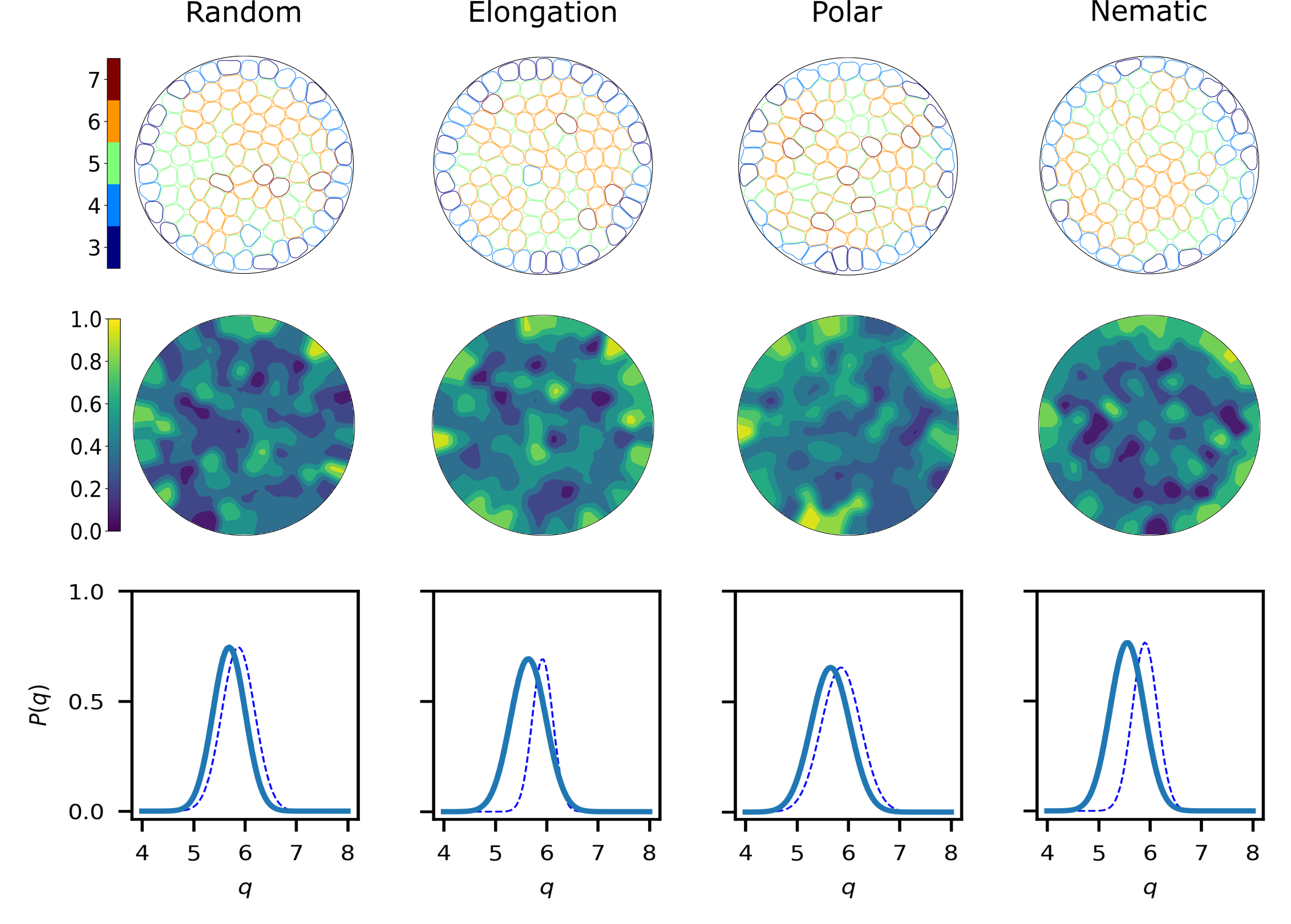}
    \caption{(first row) Cell morphology and number of neighbors. (second row) Time averaged bond number. (third row) Coordination number probability computed excluding cells in contact with confinement. The corresponding curves from Figure \ref{fig:sl1} are shown for comparison (dashed curves).}
    \label{fig:confinement}
\end{figure}

To further analyse the emerging properties in the circular confinement Figure \ref{fig:confinement} shows snapshots of the configuration, highlighting the cell morphology and their neighbour relations. We also compute the bond number $|\Psi_6|$, to be $1$ for a perfectly hexagonal arrangement and $0$ for an isolated cell, see \cite{Loewe_PRL_2020}. The quantity is computed locally and averaged over time, essentially showing a global liquid like behavior for all four models. The coordination number probability is computed as in Figure \ref{fig:sl1}, but excluding the cells in contact with the confinement. All four models show a decrease in the mean value as a result of the confinement. The elongation and nematic models also show an increase in the variance if compared with the results in Figure \ref{fig:sl1} and thus indicate a shift of the solid-to-liquid transition towards lower activities in the phase diagram.

In real systems confinement has a tremendous effect on the emerging macroscopic behaviour and might even induce morphgenesis-like processes. Our simulation results indicate that the emerging behavior in confinements strongly depends on subcellular details and the way how activity is enforced on the cellular level in the modeling approach.  

\section{Conclusions}

We use multiphase field models to analyse confluent monolayers of deformable cells. The advantage of such a modeling approach has been pointed out in various recent contributions \cite{Nonomura_PLOS_2012,Camley_PNAS_2014,Palmieri_SR_2015,Mueller_PRL_2019,Wenzel_JCP_2019,Loewe_PRL_2020}. Cell deformations and detailed cell-cell interactions, as well as subcellular details to resolve the mechanochemical interactions underlying cell migration can naturally be handled. Also topological changes, such as T1 transitions, follow naturally in a multiphase field framework. Using efficient numerics and appropriate computing power we analyse the emerging macroscopic behavior in such models and compare with known universal features of cell monolayers and epithelia tissue. We consider four different minimal models. They all follow the same methodology and only differ at microscopic details on the incorporation of activity: The random model \cite{Loewe_PRL_2020} determines the direction of motion on the single cell level by a stochastic process, the elongation model \cite{Mueller_PRL_2019} aligns the direction of motion with the long axis of the cell and two models, the polar \cite{Wenzel_JCP_2019} and a nematic model, which use subcellular details to determine strength and direction of motion on a single cell level. 

Various of the known generic features of confluent monolayers are reproduced by all four models, highlighting the robustness of these features on microscopic details. This includes solid-to-liquid transition, which leads after appropriate calibration of parameters to similar phase diagrams as obtained with vertex and voronoi models \cite{Bi_PRX_2016}. Other common features are the spontaneous formation of vortices and topological defects as well as the emergence of active turbulent flows. 

However, the four models also lead to different results if more quantitative measures are considered. This becomes apparent for the deformation of cells. While the shape variability of the cells can be described by a k-Gamma distribution over a broade range of parameters for all four models, the narrow range of the parameter $k$ found in \cite{Atia_NP_2018} for various epithelia systems, could only be reproduced by the elongation and polar model. But not only geometrical properties of the cells, also topological features differ between the four models. The ratio of multicellular rosettes depends on the microscopic details. As these rosettes provide an efficient mechanism for tissue remodeling, see e.g. \cite{Yan_PRX_2019}, these differences need to be considered in further model extensions. The most striking differences between the four model are found by analysing the emerging nematic liquid crystal properties of the monolayer and its topological defects. The different role of $+ \frac{1}{2}$ and $- \frac{1}{2}$ defects in active nematodynamics can only be reproduced by the elongation, polar and nematic model. However, the mechanical properties differ. The elongation model is constructed to produce extensile behaviour on the multicellular level. In the polar model the contractile behavior on the single cell level carries over to the multicellular level and also the nematic model, where the properties on the single cell level depend on shape, leads to contractile behavior on the multicellular level. As suggested by the experiments on MDCK cells \cite{balasubramaniamEtAl2021} the emerging mechanical properties of these models on the multicellular level might be influenced by changing the considered cell-cell interactions. Also the simulations in confinement bring differences of the models to light. Induced global rotation, as observed in circular confinements for MDCK cells in \cite{Deforet_NC_2014} and reproduced by particle-based simulations \cite{Hakim_RPP_2017} could only be observed with the polar model.  However, all models show a slight change in coordination number distribution.

The comprehensive comparison of multiphase field models for confluent cell monolayers shows the strong effect of the way activity is considered on a single cell level and highlights the need to take these effects into account for predictive simulation results at the multicellular level. However, the results also show the robustness of these models in producing generic qualitative features for cell monolayers and epithelia tissue. The flexibility of multiphase field models, not only in terms of cell deformability and topological changes, such as T1 transitions, but also in incorporating mechanochemical effects on a single cell level and for cell-cell interactions offers this modeling approach a huge potential for multiscale simulations of multicellular dynamics.

\begin{acknowledgments}
A.V. acknowledges support by the German Research Foundation (DFG) under Grant FOR3013. We further acknowledge computing resources provided at J\"ulich Supercomputing Center under Grant No. pfamdis.
\end{acknowledgments}

\bibliography{library}
\newpage
\appendix


\end{document}